# Sputtered MoRe SQUID-on-tip for high-field magnetic and thermal nanoimaging

Kousik Bagani[1,*], Jayanta Sarkar[1,§], Aviram Uri[1], Michael L. Rappaport[2], Martin E. Huber[3], Eli Zeldov[1,*], and Yuri Myasoedov[1]

[1]*Department of Condensed Matter Physics, Weizmann Institute of Science, Rehovot 7610001, Israel*
[2] *Physics Core Facilities, Weizmann Institute of Science, Rehovot 7610001, Israel*
[3]*Departments of Physics and Electrical Engineering, University of Colorado Denver, Denver, Colorado 80217, USA*

ABSTRACT:

Scanning nanoscale superconducting quantum interference devices (SQUIDs) are gaining interest as highly sensitive microscopic magnetic and thermal characterization tools of quantum and topological states of matter and devices. Here we introduce a novel technique of collimated differential-pressure magnetron sputtering for versatile self-aligned fabrication of SQUID-on-tip (SOT) nanodevices, which cannot be produced by conventional sputtering methods due to their diffusive, rather than the required directional point-source, deposition. The new technique provides access to a broad range of superconducting materials and alloys beyond the elemental superconductors employed in the existing thermal deposition methods, opening the route to greatly enhanced SOT characteristics and functionalities. Utilizing this method, we have developed MoRe SOT devices with sub-50 nm diameter, magnetic flux sensitivity of 1.2 $\mu\Phi_0/Hz^{1/2}$ up to 3 T at 4.2 K, and thermal sensitivity better than 4 $\mu K/Hz^{1/2}$ up to 5 T – about five times higher than any previous report – paving the way to nanoscale imaging of magnetic and spintronic phenomena and of dissipation mechanisms in previously inaccessible quantum states of matter.

Scanning probe microscopy based on micro- and nano-scale superconducting quantum interference devices (SQUIDs) has attracted growing attention in recent years in a broad range of fields [1–9]. The significant technological advances in the development of versatile scanning SQUID probes and imaging methods have led to key insights into microscopic properties of superconductors [4,10,11], vortex dynamics [12–16], magnetism at oxide interfaces [17–21], edge transport and magnetism in topological states of matter [22–24], and dynamics of phase transitions [25]. The two main approaches to the fabrication of scanning SQUIDs are based on planar lithographic techniques [26–33] and on self-aligned deposition on a quartz pipette, forming a SQUID-on-tip (SOT) [34–38]. The planar scanning SQUIDs have major advantages in terms of robustness, high sensitivity for imaging local magnetic fields and currents, variable sample temperature, and integration of multifunctional capabilities including local susceptibility measurements and local stress response characterization [26,29,30,39]. The scanning SOTs, on the other hand, have the benefits of nanoscale spatial resolution, high spin sensitivity, and operation at elevated magnetic fields [36]. Moreover, the SOT has recently been demonstrated to have outstanding temperature sensitivity, establishing a new tool of cryogenic thermal imaging that allows spatially resolved study of nanoscale dissipation processes in quantum systems [40,41]. The combination of record thermal and spin sensitivities renders the scanning SOT a powerful tool for investigation of a wide range of novel materials and quantum states of matter. One of the main limitations of SOTs hitherto, however, has been the restricted range of working temperatures and magnetic fields. As a concrete example, scanning SOT thermometry has been recently utilized to image work and dissipation in the quantum Hall state in graphene, revealing the microscopic mechanisms undermining dissipationless transport due to edge reconstruction [42]. Study of dissipation and of the various predicted edge structures in the fractional quantum Hall state, however, is currently unattainable with the existing SOTs due to the required high-field and low-temperature operation. Here we present a new fabrication method that establishes a versatile platform for greatly diversifying the applicable superconducting materials and thus significantly expands the range of SOT characteristics and potential functionalities. Utilizing this method we fabricated MoRe SOTs that operate in fields of up to 5 T opening the route to nanoscale magnetic studies and thermal imaging of dissipation mechanisms in previously inaccessible quantum states of matter.

In contrast to planar SQUID devices, SOTs require no lithographic or microfabrication processes. Their fabrication is based on self-aligned deposition of a superconductor onto a pulled, hollow quartz pipette using three deposition steps [34,36]. Superconducting leads are formed on the opposite sides of the pipette by two depositions with the pipette rotated by ±110° about an axis perpendicular to its length, followed by a head-on deposition of a superconducting loop at the pipette apex. In order to prevent shorting between the leads, this self-aligned method requires deposition from a point-like source, and hence heretofore the SOTs were fabricated only by physical vapor deposition, using thermal evaporation from a resistive boat or from an electron-beam source. Since thermal deposition, except in rare cases, causes fractionalization of alloys, it is usually limited to elemental materials. Therefore, SOT fabrication has been demonstrated so far only with Al [34], Pb [36–38,40,41], Nb [36], and In [23,24]. The restriction of using only elemental superconductors poses a severe limitation on the applicable superconducting materials and the corresponding SOT characteristics, including the critical temperature $T_c$ and the critical magnetic field $H_{c2}$.

Sputtering is the most common method for thin-film deposition of superconducting devices [43–47]. Applying this technique would open a new avenue for SOT fabrication from a wide range of materials and alloys, including multilayer structures, thus allowing optimization of SOT parameters for numerous applications. Conventional sputtering methods, however, are incompatible with the self-aligned SOT fabrication, which requires a point-like source and ballistic line-of-sight propagation of the deposition elements. Sputtering targets are usually large, leading to deposition from a wide range of angles, and the sputtering process requires relatively high gas pressures (∼5-10 mTorr) resulting in diffusive propagation of the elements. As a result, such deposition produces an almost isotropic coating of the pipette over its entire surface, thus shorting the superconducting electrodes (as confirmed by our early tests).

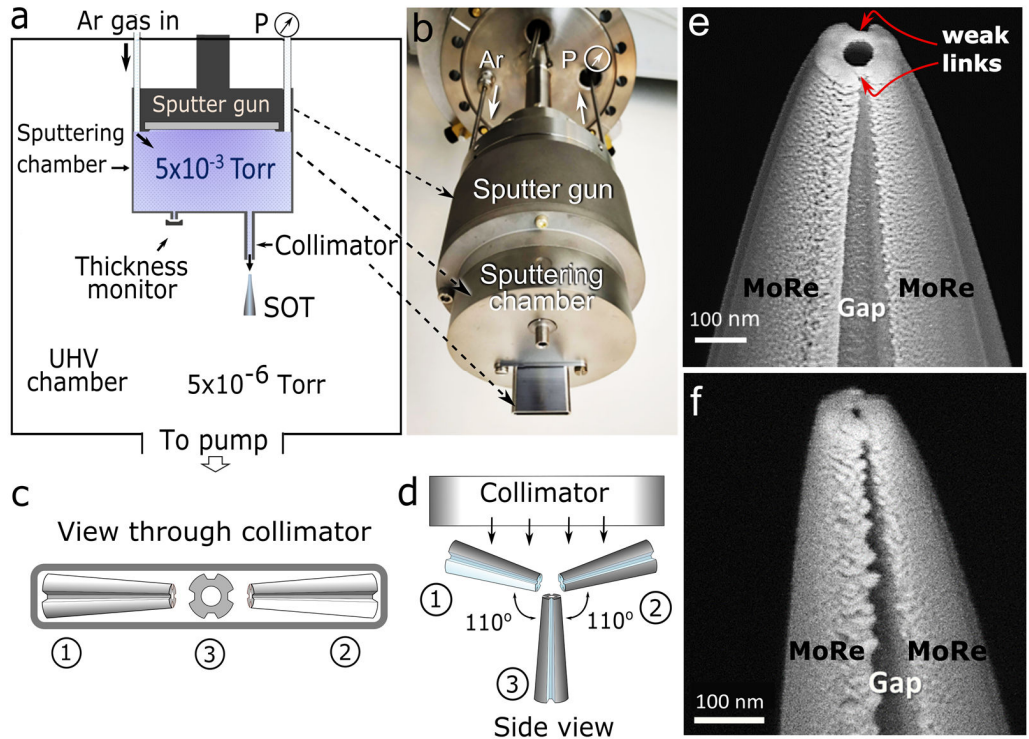

FIG. 1: (**a**) A schematic drawing of the collimated differential-pressure sputtering method. The sputter gun is sealed in a small sputtering chamber and the collimator allows the sputtered elements to deposit on the quartz pipette placed under it in the main UHV chamber. P is the pressure gauge used to measure the pressure inside the sputtering chamber. (**b**) Photograph of the sputtering gun assembly showing the sputtering chamber, collimator, and the tube for thickness monitor. (**c**) Cross section of the collimator, showing positions 1 to 3 of the quartz pipette during the three deposition steps. (**d**) Side view of the collimator showing the geometry of the three deposition steps at different pipette angles. (**e**), (**f**) Scanning electron microscope images of sputtered MoRe SOTs showing two superconducting leads with a clear gap between them and the superconducting loop at the apex with two weak links forming the SQUID. Effective loop diameters of the SQUIDs are 105 nm (**e**) and 49 nm (**f**).

Several groups have developed collimated sputtering [48–50] for deposition on large wafers using a matrix of collimation plates and lower than usual sputtering pressure of ∼$10^{-4}$ Torr to prevent diffusive scattering. To sustain the plasma at such a low pressure, however, it is necessary to complement the sputter source with an electron gun.

Here we present a novel collimated differential-pressure magnetron sputtering method for small substrates implemented by a simple addition of a "sputtering chamber" to a commercial sputtering source (AJA A320) that does not necessitate an additional electron source (Fig. 1). The Ar gas is continuously supplied to the sputtering chamber through a feed tube and an additional tube provided for monitoring the gas pressure in the chamber (Figs. 1a,b). A collimator with a racetrack cross section of ~2×20 mm² and 25 mm long (Figs. 1b,c) allows the Ar gas and the sputtered elements to flow into the cryopumped ultra-high vacuum (UHV) chamber whose base pressure is 1×10⁻⁹ Torr. The SOT deposition assembly is placed about 20 mm below the collimator (Fig. 1d). A quartz crystal thickness monitor is placed under a Ø3 mm and 5 mm long tube for monitoring the deposition rate. The flow impedances of the collimator and tube were designed to maintain a large pressure difference between the two chambers. By tuning the flow of the incoming Ar, the pressure inside the sputtering chamber was set to ~5 mTorr, while the pressure inside the UHV chamber increased only to ~5×10⁻⁶ Torr, thus eliminating the need for a throttle valve used to maintain the required high pressure in conventional sputtering systems. These pressures allow proper operation of the sputter gun on the one hand, while offering the ballistic-flow deposition from a narrow source, as required for the SOT self-aligned fabrication, on the other hand.

Utilizing this setup, we have fabricated SOT devices using a 50-50 weight percent MoRe target. A grooved quartz tube [38] was pulled to a sharp pipette, followed by three deposition steps at different tilt angles (Figs. 1c,d) that formed the two superconducting electrodes connected to a superconducting ring at the tip apex. We used the common sputtering conditions [44–47] with Ar pressure of ~5 mTorr and 300W DC power. The MoRe film thickness was 15 to 20 nm, deposited at a rate of ~1.5 Å/sec. Several MoRe SOTs with various diameters were fabricated and some of them were vacuum annealed at 700 °C for 1 hour to improve film quality. The SEM micrograph of a 105 nm diameter MoRe SOT in Fig. 1e shows a clear gap between the two electrodes. A sharp gap is well resolved even in a very small SOT with an apex diameter of 49 nm (Fig. 1f), indicating that the collimated sputtered atoms propagate ballistically. The effective SOT diameters $D$ were extracted from the magnetic field periodicity $\Delta B$ of quantum interference pattern of the SQUID critical current, $D = 2\sqrt{\Phi_0/\pi\Delta B}$, where $\Phi_0 = h/2e$ is the magnetic flux quantum, $h$ is Planck's constant and $e$ the electron charge.

The MoRe SOTs were characterized at 4.2 K using the electrical circuit in Fig. 2a. The voltage source $V_b$ combined with the large bias resistor $R_b$ act as a current source with a bias current $I_b = V_b/R_b$ while the small shunt resistor, $R_s \ll R_b$, provides an effective voltage bias to the SOT ($R_p$ is the parasitic resistance of wires and SOT contacts). In addition to the external shunt $R_s$, the SOT is resistively shunted by a gold thin film deposited onto the pipette at a distance of ~0.5 mm away from the tip apex that bridges the two superconducting leads. This on-tip shunt resistance $R_{sh}$ damps the SOT high frequency dynamics, suppressing hysteresis. The current through the SOT, $I_{SOT}$, was measured using a cryogenic SQUID series array amplifier (SSAA) [51]. Figure 2a presents the measured current-voltage characteristics $I_{SOT}(V_b)$ of the 49 nm diameter SOT (device *A*, Fig. 1f) at several values of the applied field $B_a$, while the corresponding calculated $I_{SOT}(V_{SOT})$ are shown in Fig. 2b. The $I_{SOT}(V_{SOT})$ characteristics of the SOT show overdamped behavior with critical current of $I_c = 35$ µA at zero magnetic field and the high-bias differential resistance determined by $R_{sh} \cong 5.9$ Ω. The SOT remains superconducting even at $B_a = 5$ T (the maximal field in our setup) with $I_c = 4$ µA. The critical magnetic field ($H_{c2}$) of the MoRe being over 5 T indicates that the SOT can operate over a very wide range of magnetic fields, about five times higher than in the previously reported Pb and Nb SOTs [36]. The $I_c$ is found to increase significantly with annealing, indicating improved MoRe film quality, consistent with previous studies [47,52].

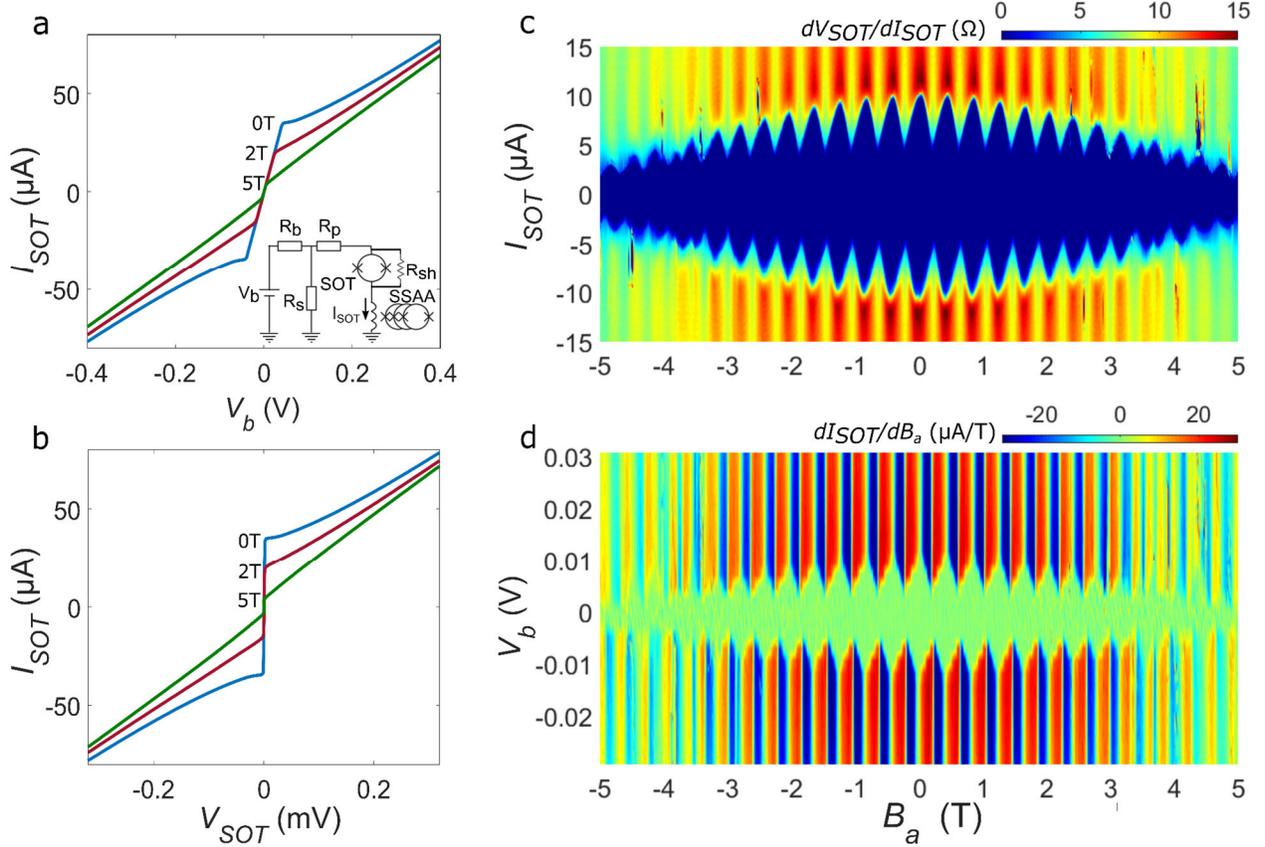

FIG. 2: Current-voltage characteristics of MoRe SOTs at 4.2 K. (**a**) Measured $I_{SOT}$ vs. $V_b$ and (**b**) the corresponding calculated $I_{SOT}$ vs. $V_{SOT}$ curves of device *A* (⌀49 nm) at different indicated magnetic fields. Inset in (a) shows the SOT measurement circuit schematic, where $R_b = 1$ kΩ, $R_s = 1$ Ω, $R_p \approx 0.5$ Ω, and $R_{sh} \approx 5$ Ω. (**c**) Differential resistance $dV_{SOT}/dI_{SOT}$ as a function of $B_a$ and $I_{SOT}$ of device *B* showing SQUID interference pattern corresponding to an effective loop diameter of 79 nm. (**d**) Magnetic response function $dI_{SOT}/dB_a$ of device *B* plotted as a function of $B_a$ and $V_b$.

Figure 2c shows the quantum interference pattern of MoRe SOT device *B* that was annealed in vacuum at 700 °C for one hour. It presents the color-rendered differential resistance $dV_{SOT}/dI_{SOT}$ vs. $B_a$ and $I_{SOT}$ derived from the measured current-voltage characteristics. The modulation period of 0.42 T corresponds to an effective SQUID loop diameter of 79 nm. A slight canting and a relative shift in the modulation patterns for positive and negative biases is observed, which indicates a small asymmetry in the critical current of the two junctions. The quantum interference patterns with prominent oscillations in $I_c$ persist up to record high magnetic field of 5 T. The patterns are smooth and periodic up to $B_a \cong 3$ T, while at higher fields the modulation becomes irregular and the modulation depth decreases.

The magnetic response function $dI_{SOT}/dB_a$ of the SOT (device *B*), presented in Fig. 2d, shows high values up to 3.5 T followed by a pronounced decrease at higher fields as the field modulation of $I_c$ decreases. The spectral density of the current noise of device *B* (Fig. 3a) exhibits $1/f$ behavior at low frequencies followed by white noise above ~200 Hz and is only weakly magnetic field dependent as demonstrated by the 0 and 3 T curves in Fig. 3a. The white noise reaches 11 pA/Hz$^{1/2}$ in device *B* and 8 pA/Hz$^{1/2}$ in device *C* (⌀105 nm, Fig. 1e). These values are comparable to the state-of-the-art Pb SOT devices [36,38,40,41]. By applying a constant bias $V_b = 0.02$ V to device *B* and measuring the white noise $S_I^{1/2}$ and the magnetic response $dI_{SOT}/dB_a$ upon sweeping the magnetic field, we derive the flux noise $S_\Phi^{1/2} = S_I^{1/2}/|dI_{SOT}/d\Phi_a|$, where $\Phi_a = AB_a$ is the applied flux in the SOT with effective area $A$. The resultant flux noise in Fig. 3b exhibits peaks and valleys corresponding to the regions of low and high magnetic response of the interference pattern, respectively. In the sensitive regions, the MoRe SOT flux noise $S_\Phi^{1/2}$ of 0.9 to 1.5 μΦ$_0$/Hz$^{1/2}$ is lower than that of previously reported Al and Nb SOTs [34,36], but significantly higher than that of Pb SOTs [36,38]. This reduced sensitivity is a result of two factors. Firstly, lower critical current $I_c$ leading to lower magnetic response, and secondly, slightly higher current noise $S_I^{1/2}$. Achieving higher critical currents should be possible by further tuning of the deposition parameters. The remarkable feature in Fig. 3b, however, is the unprecedentedly wide range of operating fields, 0 to 3 T, which is essential for the study of high-field magnetic and transport phenomena in condensed matter systems. The right-hand axis of Fig. 3b shows the corresponding spectral density of the spin noise, $S_n^{1/2}$, of the SOT in units of μ$_B$/Hz$^{1/2}$ calculated using [53] $S_n^{1/2} = S_\Phi^{1/2} r/r_e$, where $r_e$ is the classical electronic radius, assuming spins oriented perpendicular to the SQUID loop of radius $r$ and located at the center of the loop. In the sensitive regions, device $B$ shows spin sensitivity of $S_n^{1/2} = 15$ to $30$ μ$_B$/Hz$^{1/2}$ in magnetic fields of up to 3 T.

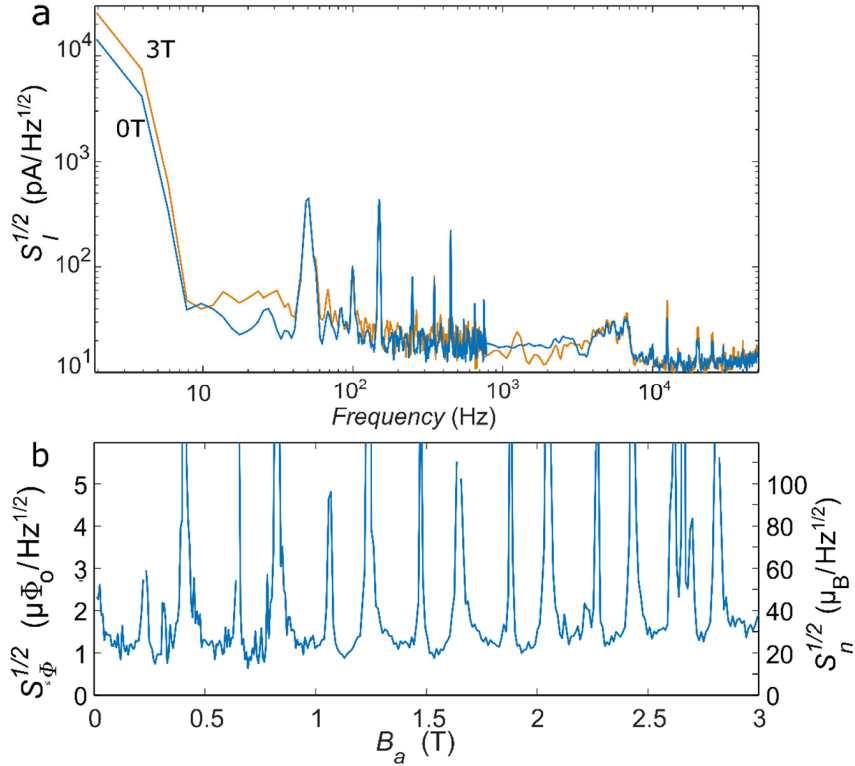

FIG. 3: Noise characteristics of SOT device $B$ (⌀79 nm) at 4.2 K. (**a**) Spectral density of the current noise $S_I^{1/2}$ at $B_a$ of 0 T and 3 T exhibits no significant change in white noise level with magnetic field. (**b**) Flux noise $S_\Phi^{1/2}$ (left axis) and spin noise $S_n^{1/2}$ (right axis) as function of magnetic field showing flux noise in the sensitive regions in the range of $S_\Phi^{1/2} = 0.9$ to $1.5$ μΦ$_0$/Hz$^{1/2}$ and spin noise of $S_n^{1/2} = 15$ to $30$ μ$_B$/Hz$^{1/2}$ up to 3 T.

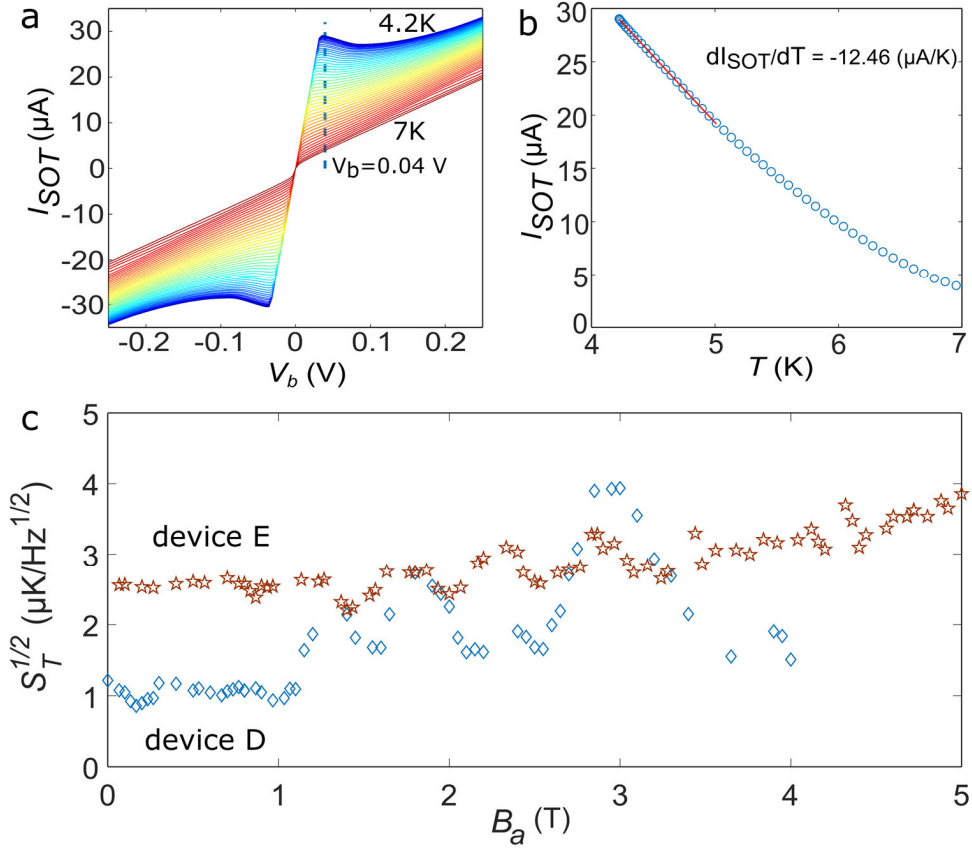

FIG. 4: Thermal response characteristics of MoRe SOTs. (**a**) $I_{SOT}$ vs. $V_b$ characteristics of device *C* measured at zero field and different temperatures ranging from 4.2 K to 7 K. (**b**) $I_{SOT}$ vs. $T$ measured at a constant bias voltage $V_b = 0.04$ V (dashed line in (a)). The SOT thermal response $dI_{SOT}/dT$ was obtained from the slope of the linear fit between 4.2 and 5 K (red). (**c**) $S_T^{1/2}$ as function of $B_a$ showing thermal noise of 1 to 2 µK/Hz$^{1/2}$ in the sensitive regions in magnetic fields up to 4 T in device *D* (⌀65 nm) and thermal noise of 2.5 to 3.8 µK/Hz$^{1/2}$ in the entire field range of up to 5 T in device *E* (⌀77 nm).

Scanning SOT microscopy has recently been applied to cryogenic nanoscale thermal imaging [40,41]. In order to characterize the thermal sensitivity of the MoRe SOTs, the current-voltage characteristics at a constant magnetic field were measured upon varying the temperature from 4.2 K to 7 K. The resulting $I_{SOT}$–$V_b$ characteristics of device *C* at zero magnetic field are shown in Fig. 4a. From these characteristics, the temperature dependence of the current $I_{SOT}(T)$ at a fixed bias $V_b = 0.04$ V was derived (Fig. 4b). The slope of the curve yields the thermal response of $dI_{SOT}/dT = -12.46$ µA/K at 4.2 K, which is essentially temperature independent up to 5 K and decreases in magnitude at higher temperatures. The white noise of the device at $V_b = 0.04$ V was measured to be $S_I^{1/2} = 12$ pA/Hz$^{1/2}$ giving rise to thermal noise of $S_T^{1/2} = S_I^{1/2}/|dI_{SOT}/dT| = 960$ nK/Hz$^{1/2}$ at $B_a = 0$ T, comparable to the reported ultra-low thermal noise of Pb SOTs [40,41]. Device *B* showed that the thermal response is essentially field independent. Measured at 0 T and 2 T the response was $dI_{SOT}/dT = -4.19$ µA/K and $-4.16$ µA/K, respectively. The current noise of this device was <11 pA/Hz$^{1/2}$ up to 3 T giving rise to thermal sensitivity better than 2.7 µK/Hz$^{1/2}$ in an unmatched field range. A more detailed study of field dependence of the thermal response was carried out on two additional SOTs (devices *D* and *E*). Since the thermal response depends on $V_b$ (see Fig. 4a), we tuned the $V_b$ to maximize the thermal response at each value of the applied field and measured the corresponding current noise. The resulting thermal noise of the SOT, $S_\text{T}^{1/2}$, as a function of magnetic field up to 5 T is presented in Fig. 4c. The thermal sensitivity of device $D$ ($I_c = 30$ µA, $\varnothing 65$ nm) reaches 1 µK/Hz$^{1/2}$ and remains essentially constant up to 1 T followed by oscillating behavior. The increase in thermal noise above 1 T is caused by a drop in the critical current in this device at higher fields accompanied by oscillations. In the more sensitive field regions, however, the thermal sensitivity remains better than 2 µK/Hz$^{1/2}$ up to 4 T. Device $E$ ($I_c = 11$ µA, $\varnothing 77$ nm) displayed thermal noise of 2.52 µK/Hz$^{1/2}$ at zero field, which gradually increases up to 3.8 µK/Hz$^{1/2}$ at 5 T.

In conclusion, we have developed a new technique of collimated differential-pressure magnetron sputtering for self-aligned fabrication of SQUID-on-tip devices. In contrast to the previous method of thermal deposition, this approach allows versatile use of a wide range of materials utilizing the well-developed standard procedures of sputtering of superconducting films, including NbN, MoGe, WSi, MgB$_2$, and possibly even high-$T_c$ superconductors. This advance is crucial for expanding the high-sensitivity nanoscale scanning magnetic and thermal microscopy techniques into new ranges of operating temperatures and magnetic fields essential for the study of magnetic phenomena and dissipation mechanisms in a wide variety of quantum systems, unconventional superconductors, and topological materials. Moreover, the developed technique has the potential for exploiting multilayer structures for new functionalities and characteristics of the SOTs. Using this novel technique we have fabricated MoRe SOTs that operate up to an unprecedentedly high magnetic field of 5 T with spin sensitivity of below 30 µ$_B$/Hz$^{1/2}$ and thermal sensitivity of better than 4 µK/Hz$^{1/2}$.

AUTHOR INFORMATION

**Corresponding Authors**

*E.Z. e-mail: eli.zeldov@weizmann.ac.il.

*K.B. e-mail: kousik.bagani@weizmann.ac.il

**Present Addresses**

§Department of Applied Physics, Aalto University, Espoo, FI-02150, Finland.

**Notes**

The authors declare no competing financial interest.

ACKNOWLEDGEMENTS

This work was supported by the European Research Council (ERC) under the European Union's Horizon 2020 research and innovation program (grant No 785971), by the US-Israel Binational Science Foundation (BSF) (grant No 2014155), by the Minerva Foundation with funding from the Federal German Ministry of Education and Research, and by the Leona M. and Harry B. Helmsley Charitable Trust grant 2018PG-ISL006. EZ acknowledges the support by the Weston Nanophysics Challenge Fund.

REFERENCES

1. M. José Martínez-Pérez and D. Koelle, ''NanoSQUIDs: Basics and recent advances'', *Phys. Sci. Rev.* **2**, 5001 (2017).

2. C. Granata and A. Vettoliere, ''Nano Superconducting Quantum Interference device: A powerful tool for nanoscale investigations'', *Phys. Rep.* **614**, 1 (2016).

3. J. R. Kirtley, L. Paulius, A. J. Rosenberg, J. C. Palmstrom, C. M. Holland, E. M. Spanton, D. Schiessl, C. L. Jermain, J. Gibbons, Y.-K.-K. Fung, M. E. Huber, D. C. Ralph, M. B. Ketchen, G. W. Gibson, and K. A. Moler, ''Scanning SQUID susceptometers with sub-micron spatial resolution'', *Rev. Sci. Instrum.* **87**, 093702 (2016).

4. J. R. Kirtley, ''Fundamental studies of superconductors using scanning magnetic imaging'', *Reports Prog. Phys.* **73**, 126501 (2010).

5. R. Kleiner, D. Koelle, F. Ludwig, and J. Clarke, ''Superconducting quantum interference devices: State of the art and applications'', *Proc. IEEE* **92**, 1534 (2004).

6. R. L. Fagaly, ''Superconducting quantum interference device instruments and applications'', *Rev. Sci. Instrum.* **77**, 101101 (2006).

7. C. Veauvy, K. Hasselbach, and D. Mailly, ''Scanning μ-superconduction quantum interference device force microscope'', *Rev. Sci. Instrum.* **73**, 3825 (2002).

8. D. Koelle, R. Kleiner, F. Ludwig, E. Dantsker, and J. Clarke, ''High-transition-temperature superconducting quantum interference devices'', *Rev. Mod. Phys.* **71**, 631 (1999).

9. J. R. Kirtley and J. P. Wikswo, ''Scanning SQUID microscopy'', *Annu. Rev. Mater. Sci.* **29**, 117 (1999).

10. J. R. Kirtley, C. C. Tsuei, Ariando, C. J. M. Verwijs, S. Harkema, and H. Hilgenkamp, ''Angle-resolved phase-sensitive determination of the in-plane gap symmetry in YBa2Cu3O7−δ'', *Nat. Phys.* **2**, 190 (2006).

11. D. J. Hykel, C. Paulsen, D. Aoki, J. R. Kirtley, and K. Hasselbach, ''Magnetic fields above the superconducting ferromagnet UCoGe'', *Phys. Rev. B* **90**, 184501 (2014).

12. T. Nishio, V. H. Dao, Q. Chen, L. F. Chibotaru, K. Kadowaki, and V. V. Moshchalkov, ''Scanning SQUID microscopy of vortex clusters in multiband superconductors'', *Phys. Rev. B* **81**, 020506 (2010).

13. N. Kokubo, S. Okayasu, A. Kanda, and B. Shinozaki, ''Scanning SQUID microscope study of vortex polygons and shells in weak-pinning disks of an amorphous superconducting film'', *Phys. Rev. B* **82**, 014501 (2010).

14. L. Embon, Y. Anahory, A. Suhov, D. Halbertal, J. Cuppens, A. Yakovenko, A. Uri, Y. Myasoedov, M. L. Rappaport, M. E. Huber, A. Gurevich, and E. Zeldov, ''Probing dynamics and pinning of single vortices in superconductors at nanometer scales'', *Sci. Rep.* **5**, 7598 (2015).

15. F. S. Wells, A. V. Pan, X. R. Wang, S. A. Fedoseev, and H. Hilgenkamp, ''Analysis of low-field isotropic vortex glass containing vortex groups in YBa2Cu3O7–x thin films visualized by scanning SQUID microscopy'', *Sci. Rep.* **5**, 8677 (2015).

16. L. Embon, Y. Anahory, Ž. L. Jelić, E. O. Lachman, Y. Myasoedov, M. E. Huber, G. P. Mikitik, A. V. Silhanek, M. V. Milošević, A. Gurevich, and E. Zeldov, ''Imaging of super-fast dynamics and flow instabilities of superconducting vortices'', *Nat. Commun.* **8**, 85 (2017).

17. J. A. Bert, B. Kalisky, C. Bell, M. Kim, Y. Hikita, H. Y. Hwang, and K. A. Moler, ''Direct imaging of the coexistence of ferromagnetism and superconductivity at the LaAlO3/SrTiO3 interface'', *Nat. Phys.* **7**, 767 (2011).

18. B. Kalisky, E. M. Spanton, H. Noad, J. R. Kirtley, K. C. Nowack, C. Bell, H. K. Sato, M. Hosoda, Y. Xie, Y. Hikita, C. Woltmann, G. Pfanzelt, R. Jany, C. Richter, H. Y. Hwang, J. Mannhart, and K. A. Moler, ''Locally enhanced conductivity due to the tetragonal domain structure in LaAlO3/SrTiO3 heterointerfaces'', *Nat. Mater.* **12**, 1091 (2013).

19. X. R. Wang, C. J. Li, W. M. Lu, T. R. Paudel, D. P. Leusink, M. Hoek, N. Poccia, A. Vailionis, T. Venkatesan, J. M. D. Coey, E. Y. Tsymbal, Ariando, and H. Hilgenkamp, ''Imaging and control of ferromagnetism in LaMnO3/SrTiO3 heterostructures'', *Science* **349**, 716 (2015).

20. Y. Anahory, L. Embon, C. J. Li, S. Banerjee, A. Meltzer, H. R. Naren, A. Yakovenko, J. Cuppens, Y. Myasoedov, M. L. Rappaport, M. E. Huber, K. Michaeli, T. Venkatesan, Ariando, and E. Zeldov, ''Emergent nanoscale superparamagnetism at oxide interfaces'', *Nat. Commun.* **7**, 12566 (2016).

21. E. Persky and B. Kalisky, ''Scanning SQUID View of Oxide Interfaces'', *Adv. Mater.* **30**, 1706653 (2018).

22. K. C. Nowack, E. M. Spanton, M. Baenninger, M. König, J. R. Kirtley, B. Kalisky, C. Ames, P. Leubner, C. Brüne, H. Buhmann, L. W. Molenkamp, D. Goldhaber-Gordon, and K. A. Moler, ''Imaging currents in HgTe quantum wells in the quantum spin Hall regime'', *Nat. Mater.* **12**, 787 (2013).

23. E. O. Lachman, A. F. Young, A. Richardella, J. Cuppens, H. R. Naren, Y. Anahory, A. Y. Meltzer, A. Kandala, S. Kempinger, Y. Myasoedov, M. E. Huber, N. Samarth, and E. Zeldov, ''Visualization of superparamagnetic dynamics in magnetic topological insulators'', *Sci. Adv.* **1**, e1500740 (2015).

24. E. O. Lachman, M. Mogi, J. Sarkar, A. Uri, K. Bagani, Y. Anahory, Y. Myasoedov, M. E. Huber, A. Tsukazaki, M. Kawasaki, Y. Tokura, and E. Zeldov, ''Observation of superparamagnetism in coexistence with quantum anomalous Hall C = ±1 and C = 0 Chern states'', *npj Quantum Mater.* **2**, 70 (2017).

25. A. Kremen, H. Khan, Y. L. Loh, T. I. Baturina, N. Trivedi, A. Frydman, and B. Kalisky, ''Imaging quantum fluctuations near criticality'', *Nat. Phys.* **14**, 1205 (2018).

26. N. C. Koshnick, M. E. Huber, J. A. Bert, C. W. Hicks, J. Large, H. Edwards, and K. A. Moler, ''A terraced scanning super conducting quantum interference device susceptometer with submicron pickup loops'', *Appl. Phys. Lett.* **93**, 243101 (2008).

27. J. R. Kirtley, M. B. Ketchen, K. G. Stawiasz, J. Z. Sun, W. J. Gallagher, S. H. Blanton, and S. J. Wind, ''High-resolution scanning SQUID microscope'', *Appl. Phys. Lett.* **66**, 1138 (1995).

28. M. E. Huber, N. C. Koshnick, H. Bluhm, L. J. Archuleta, T. Azua, P. G. Björnsson, B. W. Gardner, S. T. Halloran, E. A. Lucero, and K. A. Moler, ''Gradiometric micro-SQUID susceptometer for scanning measurements of mesoscopic samples'', *Rev. Sci. Instrum.* **79**, 053704 (2008).

29. S. Wissberg, A. Kremen, Y. Shperber, and B. Kalisky, ''Vortex configuration in the presence of local magnetic field and locally applied stress'', *Phys. C* **533**, 114 (2017).

30. J. R. Kirtley, C. C. Tsuei, K. A. Moler, V. G. Kogan, J. R. Clem, and A. J. Turberfield, ''Variable sample temperature scanning superconducting quantum interference device microscope'', *Appl. Phys. Lett.* **74**, 4011 (1999).

31. C. Granata, A. Vettoliere, R. Russo, E. Esposito, M. Russo, and B. Ruggiero, ''Supercurrent decay in nano-superconducting quantum interference devices for intrinsic magnetic flux resolution'', *Appl. Phys. Lett.* **94**, 062503 (2009).

32. Y. Shibata, S. Nomura, H. Kashiwaya, S. Kashiwaya, R. Ishiguro, and H. Takayanagi, ''Imaging of current density distributions with a Nb weak-link scanning nano-SQUID microscope'', *Sci. Rep.* **5**, 15097 (2015).

33. F. Foroughi, J.-M. Mol, T. Müller, J. R. Kirtley, K. A. Moler, and H. Bluhm, ''A micro-SQUID with dispersive readout for magnetic scanning microscopy'', *Appl. Phys. Lett.* **112**, 252601 (2018).

34. A. Finkler, Y. Segev, Y. Myasoedov, M. L. Rappaport, L. Ne'eman, D. Vasyukov, E. Zeldov, M. E. Huber, J. Martin, and A. Yacoby, ''Self-Aligned Nanoscale SQUID on a Tip'', *Nano Lett.* **10**, 1046 (2010).

35. A. Finkler, D. Vasyukov, Y. Segev, L. Ne'eman, E. O. Lachman, M. L. Rappaport, Y. Myasoedov, E.

Zeldov, and M. E. Huber, ''Scanning superconducting quantum interference device on a tip for magnetic imaging of nanoscale phenomena'', *Rev. Sci. Instrum.* **83**, 073702 (2012).

36. D. Vasyukov, Y. Anahory, L. Embon, D. Halbertal, J. Cuppens, L. Neeman, A. Finkler, Y. Segev, Y. Myasoedov, M. L. Rappaport, M. E. Huber, and E. Zeldov, ''A scanning superconducting quantum interference device with single electron spin sensitivity'', *Nat. Nanotechnol.* **8**, 639 (2013).

37. Y. Anahory, J. Reiner, L. Embon, D. Halbertal, A. Yakovenko, Y. Myasoedov, M. L. Rappaport, M. E. Huber, and E. Zeldov, ''Three-Junction SQUID-on-Tip with Tunable In-Plane and Out-of-Plane Magnetic Field Sensitivity'', *Nano Lett.* **14**, 6481 (2014).

38. A. Uri, A. Y. Meltzer, Y. Anahory, L. Embon, E. O. Lachman, D. Halbertal, N. HR, Y. Myasoedov, M. E. Huber, A. F. Young, and E. Zeldov, ''Electrically Tunable Multiterminal SQUID-on-Tip'', *Nano Lett.* **16**, 6910 (2016).

39. C. A. Watson, A. S. Gibbs, A. P. Mackenzie, C. W. Hicks, and K. A. Moler, ''Micron-scale measurements of low anisotropic strain response of local Tc in Sr2RuO4'', *Phys. Rev. B* **98**, 094521 (2018).

40. D. Halbertal, J. Cuppens, M. Ben Shalom, L. Embon, N. Shadmi, Y. Anahory, H. R. Naren, J. Sarkar, A. Uri, Y. Ronen, Y. Myasoedov, L. S. Levitov, E. Joselevich, A. K. Geim, and E. Zeldov, ''Nanoscale thermal imaging of dissipation in quantum systems'', *Nature* **539**, 407 (2016).

41. D. Halbertal, M. Ben Shalom, A. Uri, K. Bagani, A. Y. Meltzer, I. Marcus, Y. Myasoedov, J. Birkbeck, L. S. Levitov, A. K. Geim, and E. Zeldov, ''Imaging resonant dissipation from individual atomic defects in graphene'', *Science* **358**, 1303 (2017).

42. A. Marguerite, J. Birkbeck, A. Aharon-Steinberg, D. Halbertal, K. Bagani, I. Marcus, Y. Myasoedov, A. K. Geim, D. J. Perello, and E. Zeldov, ''Imaging work and dissipation in the quantum Hall state in graphene'', *arXiv:1907.08973* (Nature in press) (2019).

43. S. N. Dorenbos, E. M. Reiger, U. Perinetti, V. Zwiller, T. Zijlstra, and T. M. Klapwijk, ''Low noise superconducting single photon detectors on silicon'', *Appl. Phys. Lett.* **93**, 131101 (2008).

44. N. N. Iosad, B. D. Jackson, F. Ferro, J. R. Gao, S. N. Polyakov, P. N. Dmitriev, and T. M. Klapwijk, ''Source optimization for magnetron sputter-deposition of NbTiN tuning elements for SIS THz detectors'', *Supercond. Sci. Technol.* **12**, 736 (1999).

45. V. Singh, B. H. Schneider, S. J. Bosman, E. P. J. Merkx, and G. A. Steele, ''Molybdenum-rhenium alloy based high- Q superconducting microwave resonators'', *Appl. Phys. Lett.* **105**, 222601 (2014).

46. S. Bajt, ''Improved reflectance and stability of Mo-Si multilayers'', *Opt. Eng.* **41**, 1797 (2002).

47. M. Aziz, D. Christopher Hudson, and S. Russo, ''Molybdenum-rhenium superconducting suspended nanostructures'', *Appl. Phys. Lett.* **104**, 233102 (2014).

48. S. M. Rossnagel, D. Mikalsen, H. Kinoshita, and J. J. Cuomo, ''Collimated magnetron sputter deposition'', *J. Vac. Sci. Technol. A* **9**, 261 (1991).

49. C. C. Li, H. K. Kim, and M. Migliuolo, ''Er-doped glass ridge-waveguide amplifiers fabricated with a collimated sputter deposition technique'', *IEEE Photonics Technol. Lett.* **9**, 1223 (1997).

50. S. Maidul Haque, K. Divakar Rao, S. Tripathi, R. De, D. D. Shinde, J. S. Misal, C. Prathap, M. Kumar, T. Som, U. Deshpande, and N. K. Sahoo, ''Glancing angle deposition of SiO2 thin films using a novel collimated magnetron sputtering technique'', *Surf. Coatings Technol.* **319**, 61 (2017).

51. M. E. Huber, P. A. Neil, R. G. Benson, D. A. Burns, A. M. Corey, C. S. Flynn, Y. Kitaygorodskaya, O. Massihzadeh, J. M. Martinis, and G. C. Hilton, ''DC SQUID series array amplifiers with 120 MHz bandwidth'', *IEEE Trans. Appiled Supercond.* **11**, 1251 (2001).

52. S. M. Deambrosis, G. Keppel, V. Ramazzo, C. Roncolato, R. G. Sharma, and V. Palmieri, ''A15 superconductors: An alternative to niobium for RF cavities'', *Phys. C* **441**, 108 (2006).

53. M. B. Ketchen, D. D. Awschalom, W. J. Gallagher, A. W. Kleinsasser, R. L. Sandstrom, J. R. Rozen,

and B. Bumble, ''Design, fabrication, and performance of integrated miniature SQUID susceptometers'', *IEEE Trans. Magn.* **25**, 1212 (1989).